\documentclass[12pt]{article}%
\usepackage{amssymb}
\usepackage{amsmath}
\usepackage{amsfonts}
\usepackage{graphicx}%
\providecommand{\U}[1]{\protect\rule{.1in}{.1in}}
\newtheorem{theorem}{Theorem}
\newtheorem{acknowledgement}[theorem]{Acknowledgement}

\begin{document}
\begin{titlepage}
\vspace{.3cm} \vspace{1cm}
\begin{center}
\baselineskip=16pt \centerline{\Large\bf On Unification of Gravity and Gauge Interactions} \vspace{2truecm} \centerline{\large\bf Ali H.
Chamseddine$^{1,3}$\ , \ Viatcheslav Mukhanov$^{2}$\ \ } \vspace{.5truecm}
\emph{\centerline{$^{1}$Physics Department, American University of Beirut, Lebanon}}
\emph{\centerline{$^{2}$Theoretical Physics, Ludwig Maxmillians University,Theresienstr. 37, 80333 Munich, Germany }}
\emph{\centerline{$^{3}$I.H.E.S. F-91440 Bures-sur-Yvette, France}}
\end{center}
\vspace{2cm}
\begin{center}
{\bf Abstract}
\end{center}
Considering a higher dimensional Lorentz group as the tangent symmetry, we unify gravity
and gauge interactions in a natural way. The spin connection of the gauged Lorentz group is  then responsible for
both gravity and gauge fields, and the action for the gauged fields becomes  part of the spin curvature squared.
The realistic group which unifies all known particles and interactions is the $SO(1,13)$ Lorentz group whose gauge part
leads to $SO(10)$ grand unified theory and contains double the number of required fermions in the fundamental spinor
representation. Mirror fermions could acquire mass utilizing a mechanism employed for topological superconductors.
Family unification could be achieved by considering the $SO(1,21)$ Lorentz group.
\end{titlepage}

\section{Introduction}

In General Relativity the Lorentz group is realized as a local symmetry of the
tangent manifold. There exists no spinor representations of the
diffeomorphisms and this dictates the use of this local symmetry in curved
space-time. Usually the dimension of the tangent group is taken to be equal to
the dimension of the curved manifold and the Lorentz symmetry is then simply a
manifestation of the equivalence principle for spaces without torsion.
Considering the group of local Lorentz transformations in tangent space, we
can reformulate General Relativity as a gauge theory where the gauge fields
are the spin-connections. If the dimensions of space-time and tangent space
are the same, the gauge fields (spin-connections) simply encode the same
amount of information about dynamics of the gravitational field as the affine
connections and nothing more. However, the dimension of the tangent space must
not necessarily be the same as the dimension of the manifold \cite{Wein}. In
\cite{CMtangent} we have shown that the metricity condition have unambiguous
solution also in the case when the tangent group of 4d manifold is ten
dimensional and corresponds to the de Sitter or anti de Sitter group. In such
case the theory is also completely equivalent to General Relativity. In this
paper we consider the dimension of the tangent group to be more than six and
show that this allows us to unify Yang-Mills gauge theories with gravity in
terms of higher dimensional gauged Lorentz groups. The gauge transformations
are then realized as subgroup of the tangent Lorentz group and the spinors
describing matter are \textquotedblleft unified\textquotedblright\ all being
in the fundamental representation of this higher dimensional Lorentz group.
The realistic group which unifies all particles within one family is
$SO(1,13)$ and naturally leads to Einstein gravity with the $SO\left(
10\right)  $ gauge group being, however, not entirely equivalent to the
$SO\left(  10\right)  $ grand unified theory. This, however, suffers from the
presence of mirror fermions. This problem, may be cured by considering a more
complicated model such as $SO(1,21)$ giving rise to an $SO\left(  18\right)  $
grand unified theory which utilizes topological superconductors \cite{Wen},
\cite{Kitaev} to give masses to the mirror fermions leaving only three
massless families with $SU\left(  5\right)  $ symmetry \cite{Zee}.

\section{Tangent group}

Let us consider a $4$-dimensional manifold and assume that at every point of
this manifold there is real $N$-dimensional "tangent space"\footnote{By abuse
of notaton, we refer to the space of the tangent group as the tangent space.}
spanned by linearly independent vectors $\mathbf{v}_{A}$, where $A=1,2...N.$
Assuming that $N\geq4$, the coordinate basis vectors $\mathbf{e}_{\alpha
}\equiv\partial/\partial x^{\alpha},$ where $\alpha=1,...4,$ span
$4$-dimensional (sub)space in this space. Next we define the scalar product in
the "tangent space" and take vectors $\mathbf{v}_{A}$ to be orthonormal with
respect to the \textquotedblleft Minkowski matrix\textquotedblright%
\ $\eta_{AB}$ $(-,+,...,+)$
\begin{equation}
\mathbf{v}_{A}\cdot\mathbf{v}_{B}=\eta_{AB}. \label{1}%
\end{equation}
The Lorentz transformations
\begin{equation}
\mathbf{\tilde{v}}_{A}=\Lambda_{A}^{\quad B}\mathbf{v}_{B},\text{
\ \ \ \ \ \ }\Lambda_{A}^{\quad C}\eta_{CD}\Lambda_{A}^{\quad D}=\eta
_{AB}\text{\ } \label{1a}%
\end{equation}
preserve the orthogonality of the basis vectors $\mathbf{v}_{A}$ ,
$\mathbf{\tilde{v}}_{A}\cdot\mathbf{\tilde{v}}_{B}=\eta_{AB}.$ The scalar
product of coordinate basis vectors, which also reside in the tangent space,
induces the metric in the $4$-dimensional manifold%
\begin{equation}
\mathbf{e}_{\alpha}\cdot\mathbf{e}_{\beta}=g_{\alpha\beta}(x^{\gamma}).
\label{2}%
\end{equation}
Expanding $\mathbf{e}_{\alpha}$ in $\mathbf{v}_{A}$-basis we have%
\begin{equation}
\mathbf{e}_{\alpha}=e_{\alpha}^{A}\mathbf{v}_{A}, \label{3}%
\end{equation}
where the coefficients of the expansion $e_{\alpha}^{A}$ are the vielbiens (or
soldering forms). Substituting in (\ref{2}) we obtain the following expression
for the metric $g_{\alpha\beta}$
\begin{equation}
g_{\alpha\beta}=e_{\alpha}^{A}e_{\beta}^{B}\eta_{AB}=e_{\alpha}^{A}e_{A\beta}.
\label{4}%
\end{equation}
Hereafter, we always raise and lower tangent space indices with Minkowski
metric $\eta_{AB}$. Next we consider parallel transport on the manifold
relating\ vectors in the \textquotedblleft nearby\textquotedblright\ tangent
spaces. The affine and spin-connections determining the rules of parallel
transport for coordinate basis vectors and vielbiens are defined by%
\begin{equation}
\mathbf{\nabla}_{\mathbf{e}_{\beta}}\mathbf{e}_{\alpha}\equiv\mathbf{\nabla
}_{\beta}\mathbf{e}_{\alpha}=\Gamma_{\alpha\beta}^{\nu}\mathbf{e}_{\nu
},\ \ \mathbf{\nabla}_{\beta}\mathbf{v}_{A}=-\omega_{\beta A}^{\quad
\ B}\mathbf{v}_{B}, \label{5a}%
\end{equation}
where $\mathbf{\nabla}_{\beta}$ is the derivative along a coordinate basis
vector $\mathbf{e}_{\beta}$. For example when $\mathbf{\nabla}_{\beta}$ is
applied to a scalar function $f$ it gives $\mathbf{\nabla}_{\beta}f=\partial
f/\partial x^{\beta}$. Notice that $\eta_{AB}$ and $g_{\alpha\beta}$ as
defined in (\ref{1}) and (\ref{2}) must be considered as the sets of scalar
functions and, hence, $\mathbf{\nabla}_{\beta}\eta_{AB}=0,$ $\mathbf{\nabla
}_{\gamma}g_{\alpha\beta}=\partial g_{\alpha\beta}/\partial x^{\gamma}%
\equiv\partial_{\gamma}g_{\alpha\beta}\footnote{We use the notations and
methods of Misner, Thorne, Wheeler \cite{MTW}, in particular, Chapters 9 and
10.}$.

Given $\eta_{AB},$ $g_{\alpha\beta}$ and $e_{\alpha}^{A}$ let us derive the
consistency (metricity) conditions for the connections. Taking derivative of
equation (\ref{1}) and using (\ref{5a}) we obtain%
\begin{equation}
\left(  \mathbf{\nabla}_{\alpha}\mathbf{v}_{A}\right)  \cdot\mathbf{v}%
_{B}+\mathbf{v}_{A}\cdot\left(  \mathbf{\nabla}_{\alpha}\mathbf{v}_{B}\right)
=-\omega_{\beta AB}^{\quad}-\omega_{\beta BA}^{\quad}=\mathbf{\nabla}_{\alpha
}\eta_{AB}=0, \label{6}%
\end{equation}
i.e., the spin-connection should be antisymmetric in tangent indices,
$\omega_{\beta AB}^{\quad}=-\omega_{\beta BA}^{\quad}$. Applying
$\mathbf{\nabla}_{\beta}$ to
\begin{equation}
e_{A\alpha}=\left(  \mathbf{v}_{A}\cdot\mathbf{e}_{\alpha}\right)  ,
\label{4a}%
\end{equation}
one gets%
\begin{equation}
\partial_{\beta}e_{A\alpha}=\left(  \mathbf{\nabla}_{\beta}\mathbf{v}%
_{A}\right)  \cdot\mathbf{e}_{\alpha}+\mathbf{v}_{A}\cdot\left(
\mathbf{\nabla}_{\alpha}\mathbf{e}_{\alpha}\right)  , \label{4b}%
\end{equation}
or using definitions in $\left(  \ref{5a}\right)  $%
\begin{equation}
\partial_{\beta}e_{A\alpha}=-\omega_{\beta A}^{\quad\ B}e_{B\alpha}%
+\Gamma_{\alpha\beta}^{\nu}e_{A\nu}. \label{4c}%
\end{equation}
Hereafter we assume that the space-time is torsion-free, that is,
$\Gamma_{\alpha\beta}^{\nu}=\Gamma_{\beta\alpha}^{\nu}.$ In this case, $16N$
equations $\left(  \ref{4c}\right)  $ can be solved to express $40$ affine
connections $\Gamma_{\alpha\beta}^{\nu}$ and $2N\left(  N-1\right)  $
spin-connections $\omega_{\beta AB}^{\quad\ }$ $\ $\ in terms of the
derivatives of the soldering forms $\partial_{\beta}e_{A\alpha}.$ The number
of equations matches the number of connections to be determined only if the
dimension of the tangent space is equal either to $N=4$ or $N=5$
\cite{CMtangent}. For $N\geqslant6$ the number of equation in $\left(
\ref{4c}\right)  $ is less than the number of unknown connections and
$2N^{2}-18N+40=2\left(  N-4\right)  \left(  N-5\right)  $ variables remain
undetermined by soldering forms. Let $N=n+4$, then the number of unconstrained
components of the spin-connections $\omega_{\beta A}^{\quad\ B}$ is $2n\left(
n-1\right)  $ which matches the number of $SO(n)$ gauge fields. As we will see
this allows us to account for the gauge transformations which become unified
with gravity for higher dimensional gauged Lorentz group of the tangent space.
Considering%
\begin{equation}
\partial_{\gamma}g_{\alpha\beta}=\partial_{\gamma}\left(  e_{\alpha}%
^{A}e_{A\beta}\right)  =\left(  \partial_{\gamma}e_{\alpha}^{A}\right)
e_{A\beta}+e_{\alpha}^{A}\left(  \partial_{\gamma}e_{A\beta}\right)  ,
\label{5}%
\end{equation}
and substituting in the right hand side the expression for $\partial_{\gamma
}e_{\alpha}^{A}$ from $\left(  \ref{4c}\right)  $ we find
\begin{equation}
\Gamma_{\alpha\gamma}^{\nu}g_{\nu\beta}+\Gamma_{\beta\gamma}^{\nu}g_{\alpha
\nu}=\partial_{\gamma}g_{\alpha\beta}. \label{7}%
\end{equation}
In the absence of torsion, $\Gamma_{\alpha\beta}^{\nu}=\Gamma_{\beta\alpha
}^{\nu}$, these equations are solved unambiguously, to give the well known
Christoffel connection%
\begin{equation}
\Gamma_{\alpha\beta}^{\gamma}=\frac{1}{2}g^{\gamma\sigma}\left(
g_{\alpha\sigma,\beta}+g_{\sigma\beta,\alpha}-g_{\alpha\beta,\sigma}\right)  ,
\label{8}%
\end{equation}
where $g^{\gamma\sigma}$ is inverse to $g_{\alpha\beta},$ that is,
$g^{\alpha\sigma}g_{\sigma\beta}=\delta_{\beta}^{\alpha}.$ We would like to
stress that the affine connections are determined unambiguously irrespective
of the dimension of the tangent space.

For constructing gauge invariant Lagrangians we will also need $e_{A}^{\alpha
}$ defined as%
\begin{equation}
e_{A}^{\alpha}=g^{\alpha\gamma}e_{A\gamma}, \label{10}%
\end{equation}
which can be easily seen to satisfy the metricity condition
\begin{equation}
\partial_{\beta}e_{A}^{\alpha}=-\omega_{\beta A}^{\quad\ B}e_{B}^{\alpha
}-\Gamma_{\beta\nu}^{\alpha}e_{A}^{\nu}. \label{4d}%
\end{equation}
The soldering form $e_{A}^{\alpha}$ is inverse to $e_{\beta}^{B}$ only if the
number of dimensions of the tangent space and manifold match each other. The
contraction over the tangent space indices gives%
\begin{equation}
e_{A}^{\alpha}e_{\beta}^{A}=g^{\alpha\gamma}e_{A\gamma}e_{\beta}^{A}%
=g^{\alpha\gamma}g_{\gamma\beta}=\delta_{\beta}^{\alpha}, \label{12}%
\end{equation}
however, $e_{A}^{\alpha}e_{\alpha}^{B}\neq$ $\delta_{B}^{A}.$ To prove this,
let us introduce $N-4$ orthonormal vectors $\mathbf{n}_{\hat{J}}$ orthogonal
to the subspace spanned by $\mathbf{e}_{\alpha}$, that is, $\mathbf{n}%
_{\hat{J}}\cdot\mathbf{e}_{\alpha}=0$ and $\ \mathbf{n}_{\hat{J}}%
\cdot\mathbf{n}_{\hat{I}}\mathbf{=}\delta_{\hat{J}\hat{I}},$ where $\hat
{J},\hat{I}=5,6,...,N$. The vectors $\mathbf{n}_{\hat{J}}$, $\mathbf{e}%
_{\alpha}$ form a complete basis in tangent space and therefore $\mathbf{v}%
_{A}$ can be expanded as%
\begin{equation}
\mathbf{v}_{A}=v_{A}^{\alpha}\mathbf{e}_{\alpha}+n_{A}^{\hat{J}}%
\mathbf{n}_{\hat{J}}\mathbf{.} \label{13}%
\end{equation}
Taking into account (\ref{4a}) we have%
\begin{equation}
e_{A\gamma}=\left(  \mathbf{v}_{A}\cdot\mathbf{e}_{\gamma}\right)
=v_{A}^{\alpha}g_{\alpha\gamma}, \label{14}%
\end{equation}
and hence, $v_{A}^{\alpha}=g^{\alpha\gamma}e_{A\gamma}=e_{A}^{\alpha},$ that
is, the coefficients $v_{A}^{\alpha}$ in (\ref{13}) coincide with soldering
form $e_{A}^{\alpha}$. Taking this into account one gets
\begin{equation}
\eta_{AB}=\mathbf{v}_{A}\cdot\mathbf{v}_{B}=v_{A}^{\alpha}v_{B}^{\beta
}g_{\alpha\beta}+n_{A}^{\tilde{J}}n_{\tilde{J}B}=e_{A}^{\alpha}e_{\alpha
B}+n_{A}^{\hat{J}}n_{\hat{J}B}, \label{15}%
\end{equation}
or after raising the tangent space index $B$ we obtain%
\begin{equation}
e_{A}^{\alpha}e_{\alpha}^{B}=\delta_{A}^{B}-n_{A}^{\hat{J}}n_{\hat{J}}%
^{B}\equiv P_{B}^{A} \label{16}%
\end{equation}
where $P_{B}^{A}$ is a projection operator: $P_{C}^{A}P_{B}^{C}=P_{B}^{A}.$
The components $n_{A}^{\hat{J}}$ satisfy the following relations%
\begin{equation}
n_{\hat{J}}^{A}e_{A}^{\alpha}=0,\text{ \ }n_{\hat{J}}^{A}n_{A}^{\hat{I}%
}=\delta_{\hat{J}}^{\hat{I}}. \label{16b}%
\end{equation}
To verify these relations let us consider the expansion
\begin{equation}
\mathbf{n}_{\hat{J}}=l_{\hat{J}}^{B}\mathbf{v}_{B} \label{16c}%
\end{equation}
Substituting this expression into $n_{\hat{J}A}=\left(  \mathbf{n}_{\hat{J}%
}\cdot\mathbf{v}_{A}\right)  $ we obtain $n_{\hat{J}A}=l_{\hat{J}}^{B}%
\eta_{BA}$ and hence $l_{\hat{J}}^{B}=n_{\hat{J}}^{B};$ therefore%
\begin{equation}
\mathbf{n}_{\hat{J}}=n_{\hat{J}}^{B}\mathbf{v}_{B}=n_{\hat{J}}^{B}\left(
e_{B}^{\alpha}\mathbf{e}_{\alpha}+n_{B}^{\hat{I}}\mathbf{n}_{\hat{I}}\right)
, \label{16d}%
\end{equation}
from which $\left(  \ref{16b}\right)  $ follows immediately.

In vielbiens formalism the soldering form $e_{A}^{\alpha}$ is a fundamental
quantity which is required to be invariant under the group of \textit{local}
Lorentz transformations (\ref{1a}), where $\Lambda_{A}^{\quad B}=\Lambda
_{A}^{\quad B}\left(  x\right)  .$ Under Lorentz transformations the basis
vectors $\mathbf{v}_{A}$ transform as
\begin{equation}
\mathbf{v}_{A}\rightarrow\mathbf{\tilde{v}}_{A}=\Lambda_{A}^{\quad
B}\mathbf{v}_{B}, \label{17}%
\end{equation}
and correspondingly%
\begin{equation}
\mathbf{e}_{\alpha}=e_{\alpha}^{B}\mathbf{v}_{B}=e_{\alpha}^{B}\left(
\Lambda^{-1}\right)  _{B}^{\quad A}\mathbf{\tilde{v}}_{A}=\tilde{e}_{\alpha
}^{A}\mathbf{\tilde{v}}_{A}. \label{18}%
\end{equation}
It then follows that
\begin{equation}
e_{\alpha}^{A}\rightarrow\tilde{e}_{\alpha}^{A}=e_{\alpha}^{B}\left(
\Lambda^{-1}\right)  _{B}^{\quad A},\quad\text{ }e_{A}^{\alpha}\rightarrow
\tilde{e}_{A}^{\alpha}=\Lambda_{A}^{\quad B}e_{B}^{\alpha}. \label{19}%
\end{equation}
The transformation law for the spin-connection follows from its definition:%
\begin{equation}
\tilde{\omega}_{\beta A}^{\quad\ B}\mathbf{\tilde{v}}_{B}=-\mathbf{\nabla
}_{\beta}\mathbf{\tilde{v}}_{A} \label{19a}%
\end{equation}
Substituting $\mathbf{\tilde{v}}_{B}=\Lambda_{B}^{\quad C}\mathbf{v}_{C}$ and
taking into account (\ref{5a}) we deduce that
\begin{equation}
\omega_{\beta A}^{\quad B}\rightarrow\tilde{\omega}_{\beta A}^{\quad
\ B}=\left(  \Lambda\omega_{\beta}\Lambda^{-1}\right)  _{A}^{\quad B}+\left(
\Lambda\partial_{\beta}\Lambda^{-1}\right)  _{A}^{\quad B}, \label{20}%
\end{equation}
where $\Lambda$ and $\Lambda^{-1}$ are the matrices corresponding to Lorentz
transformation and its inverse.

\section{Curvature}

To introduce the curvature for the spin-connection, consider the spinors
$\psi$ which transform in tangent space according to
\begin{equation}
\psi\rightarrow\exp\left(  \frac{1}{4}\lambda^{AB}\Gamma_{AB}\right)  \psi,
\label{21}%
\end{equation}
where $\Gamma_{AB}=\frac{1}{2}\left(  \Gamma_{A}\Gamma_{B}-\Gamma_{B}%
\Gamma_{A}\right)  $ are generators of the Lie algebra in the spinor
representation and $\Gamma_{A}$ are $N$ Dirac matrices satisfying%
\begin{equation}
\left\{  \Gamma^{A},\Gamma^{B}\right\}  =2\eta^{AB},\quad\Gamma^{\dagger
A}=\Gamma^{0}\Gamma^{A}\Gamma^{0}. \label{22}%
\end{equation}
The Dirac action
\begin{equation}%
{\displaystyle\int}
d^{4}x\sqrt{g}\,\overline{\psi}i\Gamma^{C}e_{C}^{\alpha}D_{\alpha}\psi,
\label{23}%
\end{equation}
where \
\begin{equation}
D_{\alpha}\equiv\partial_{\alpha}+\frac{1}{4}\omega_{\alpha}^{AB}\Gamma_{AB}
\label{24}%
\end{equation}
is invariant under transformations (\ref{19}), (\ref{20}) and (\ref{21}).
Notice that hermiticity of the Dirac action in (\ref{23}) is guaranteed by the
metricity condition (\ref{4c}).

Next construct the spin-connection curvature by considering the commutator of
Dirac operators%

\begin{equation}
\left[  D_{\alpha},D_{\beta}\right]  =\frac{1}{4}R_{\alpha\beta}^{\quad
AB}\Gamma_{AB}, \label{25}%
\end{equation}
where%
\begin{equation}
R_{\alpha\beta}^{\quad AB}\left(  \omega\right)  =\partial_{\alpha}%
\omega_{\beta}^{\quad AB}-\partial_{\beta}\omega_{\alpha}^{\quad AB}%
+\omega_{\alpha}^{\quad AC}\omega_{\beta C}^{\quad B}-\omega_{\beta}^{\quad
AC}\omega_{\alpha C}^{\quad B}. \label{26}%
\end{equation}
Under Lorentz transformations this spin curvature transforms as%
\begin{equation}
\left(  R_{\mu\nu}\right)  _{A}^{\quad B}\rightarrow\left(  \Lambda
R\Lambda^{\,-1}\right)  _{A}^{\quad B}. \label{27}%
\end{equation}
To relate the spin-connection curvature to the affine connection curvature
consider the identity%
\begin{equation}
\partial_{\beta}\partial_{\alpha}e_{A\gamma}-\partial_{\alpha}\partial_{\beta
}e_{A\gamma}=0. \label{28}%
\end{equation}
Substituting here the expression for $\partial e$ from $\left(  \ref{4c}%
\right)  $ and using this metricity condition one more time to express
$\partial e$ which appear after taking the derivative, we immediately arrive
at the following relation%
\begin{equation}
R_{\alpha\beta}^{\quad AB}\left(  \omega\right)  e_{B\gamma}=R_{\,\,\,\gamma
\alpha\beta}^{\rho}\left(  \Gamma\right)  e_{\rho}^{A}, \label{29}%
\end{equation}
where
\begin{equation}
R_{\,\,\,\gamma\alpha\beta}^{\rho}\left(  \Gamma\right)  =\partial_{\alpha
}\Gamma_{\beta\gamma}^{\rho}-\partial_{\beta}\Gamma_{\alpha\gamma}^{\rho
}+\Gamma_{\alpha\kappa}^{\rho}\Gamma_{\beta\gamma}^{\kappa}-\Gamma
_{\beta\kappa}^{\rho}\Gamma_{\alpha\gamma}^{\kappa}, \label{30}%
\end{equation}
is the Riemann curvature. Taking $\left(  \ref{12}\right)  $ into account, we
can express the 4d Riemann curvature from $\left(  \ref{29}\right)  $ in terms
of $R_{\alpha\beta}^{\quad AB}\left(  \omega\right)  $ as
\begin{equation}
R_{\,\,\,\gamma\alpha\beta}^{\sigma}\left(  \Gamma\right)  =e_{A}^{\sigma
}R_{\alpha\beta}^{\quad AB}\left(  \omega\right)  e_{B\gamma} \label{31}%
\end{equation}
irrespective of the number of dimensions of the tangent space. Inversely we
can express $R_{\alpha\beta}^{\quad AB}\left(  \omega\right)  $ in terms of
$R_{\,\,\,\gamma\alpha\beta}^{\sigma}\left(  \Gamma\right)  $ by using
$\left(  \ref{16}\right)  $ to obtain%
\begin{equation}
R_{\alpha\beta}^{\quad AB}\left(  \omega\right)  =R_{\alpha\beta}^{\quad
AC}\left(  \omega\right)  n_{C}^{\hat{I}}n_{\hat{I}}^{B}+R_{\,\,\,\gamma
\alpha\beta}^{\rho}\left(  \Gamma\right)  e_{\rho}^{A}e^{B\gamma}. \label{32}%
\end{equation}
Next we will show that the first term on the right hand side of this equation
can be entirely expressed in terms of the spin-connections defining the
parallel transport of vectors $\mathbf{n}_{\hat{J}}$ in the subspace of
tangent space orthogonal to those part spanned by the four coordinate basis
vectors $\mathbf{e}_{\alpha}.$ These connections, which we denote by
$A_{\beta\hat{J}}^{\quad\ \hat{I}}$ for the reasons which will become clear
later, are defined as%
\begin{equation}
\mathbf{\nabla}_{\alpha}\mathbf{n}_{\hat{J}}=-A_{\alpha\hat{J}}^{\quad
\ \hat{I}}\mathbf{n}_{\hat{I}}+B_{\alpha\hat{J}}^{\quad\ \beta}\mathbf{e}%
_{\beta} \label{33}%
\end{equation}
where indices $\hat{J}$ and $\hat{I}$ run over values $5,6,...,N.$ These
indices are also raised and lowered with the Minkowski metric $\eta_{\hat
{I}\hat{J}}$ . We now show that $B_{\alpha\hat{J}}^{\quad\ \beta}=0$ and
derive the metricity conditions for $A_{\alpha\hat{J}}^{\quad\ \hat{I}}.$ On
one hand%
\begin{equation}
\mathbf{\nabla}_{\alpha}\mathbf{v}_{A}=-\omega_{\alpha A}^{\quad\ B}%
\mathbf{v}_{B}=-\omega_{\alpha A}^{\quad\ B}\left(  e_{B}^{\gamma}%
\mathbf{e}_{\gamma}+n_{B}^{\hat{I}}\mathbf{n}_{\hat{I}}\right)  , \label{34}%
\end{equation}
where we have used $\left(  \ref{13}\right)  $ in the last equality, while on
the other hand%
\begin{align}
\mathbf{\nabla}_{\alpha}\mathbf{v}_{A}  &  =\mathbf{\nabla}_{\alpha}\left(
e_{A}^{\gamma}\mathbf{e}_{\gamma}+n_{A}^{\hat{I}}\mathbf{n}_{\hat{I}}\right)
=\left(  \partial_{\alpha}e_{A}^{\gamma}+e_{A}^{\beta}\Gamma_{\alpha\beta
}^{\gamma}\right)  \mathbf{e}_{\gamma}\nonumber\\
&  +\left(  \partial_{\alpha}n_{A}^{\hat{I}}-n_{A}^{\hat{J}}A_{\alpha\hat{J}%
}^{\quad\ \hat{I}}\right)  \mathbf{n}_{\hat{I}}+n_{A}^{\hat{I}}B_{\alpha
\hat{I}}^{\quad\ \beta}\mathbf{e}_{\beta} \label{35a}%
\end{align}
Using $\left(  \ref{34}\right)  $, $\left(  \ref{35a}\right)  $ and $\left(
\ref{4d}\right)  $ we deduce that%
\begin{equation}
B_{\alpha\hat{I}}^{\quad\ \beta}=0. \label{36}%
\end{equation}
Thus, the affine connection of the vector $\mathbf{n}_{\hat{J}}$ lies entirely
in the subspace spanned by the basis vectors $\mathbf{n}_{\hat{J}}.$ Moreover,
as it follows from $\left(  \ref{34}\right)  $ and $\left(  \ref{35a}\right)
$ that
\begin{equation}
\partial_{\alpha}n_{A}^{\hat{I}}=n_{A}^{\hat{J}}A_{\alpha\hat{J}}^{\quad
\ \hat{I}}-\omega_{\alpha A}^{\quad\ B}n_{B}^{\hat{I}}. \label{37}%
\end{equation}
Next let us define
\begin{equation}
D_{\alpha}\left(  \omega\right)  n_{A}^{\hat{I}}\equiv\partial_{\alpha}%
n_{A}^{\hat{I}}+\omega_{\alpha A}^{\quad\ C}n_{C}^{\hat{I}}, \label{37a}%
\end{equation}
and consider the commutator%
\begin{equation}
\left[  D_{\alpha}\left(  \omega\right)  ,D_{\beta}\left(  \omega\right)
\right]  n_{A}^{\hat{I}}=R_{\alpha\beta A}^{\quad\quad C}\left(
\omega\right)  n_{C}^{\hat{I}} \label{37b}%
\end{equation}
On the other hand according to $\left(  \ref{37}\right)  $%
\begin{equation}
D_{\alpha}\left(  \omega\right)  n_{A}^{\hat{I}}=n_{A}^{\hat{J}}A_{\alpha
\hat{J}}^{\quad\ \hat{I}} \label{37c}%
\end{equation}
and therefore
\begin{equation}
\left[  D_{\alpha}\left(  \omega\right)  ,D_{\beta}\left(  \omega\right)
\right]  n_{A}^{\hat{I}}=D_{\alpha}\left(  \omega\right)  \left(  n_{A}%
^{\hat{J}}A_{\beta\hat{J}}^{\quad\ \hat{I}}\right)  -\left(  \alpha
\leftrightarrow\beta\right)  =n_{A}^{\hat{J}}F_{\alpha\beta\hat{J}}%
^{\quad\quad\hat{I}}\left(  A\right)  , \label{37d}%
\end{equation}
where%
\begin{equation}
F_{\alpha\beta}^{\quad\hat{I}\hat{J}}\left(  A\right)  =\partial_{\alpha
}A_{\beta}^{\hat{I}\hat{J}}-\partial_{\beta}A_{\alpha}^{\hat{I}\hat{J}%
}+A_{\alpha}^{\hat{I}\hat{L}}A_{\beta\hat{L}}^{\quad\hat{J}}-A_{\beta}%
^{\hat{I}\hat{L}}A_{\alpha\hat{L}}^{\quad\hat{J}}. \label{39}%
\end{equation}
Thus comparing $\left(  \ref{37d}\right)  $ and $\left(  \ref{37b}\right)  $
we conclude that
\begin{equation}
R_{\alpha\beta A}^{\quad\quad C}\left(  \omega\right)  n_{C}^{\hat{I}}%
=n_{A}^{\hat{J}}F_{\alpha\beta\hat{J}}^{\quad\quad\hat{I}}\left(  A\right)
\label{39a}%
\end{equation}
and using this result in $\left(  \ref{32}\right)  $ we finally obtain
\begin{equation}
R_{\alpha\beta}^{\quad AB}\left(  \omega\right)  =F_{\alpha\beta}^{\quad
\hat{J}\hat{I}}\left(  A\right)  n_{\hat{J}}^{A}n_{\hat{I}}^{B}%
+R_{\,\,\,\gamma\alpha\beta}^{\rho}\left(  \Gamma\right)  e_{\rho}%
^{A}e^{B\gamma}. \label{40}%
\end{equation}
To get the Lagrangian for the theory we have to build curvature invariants out
of $R_{\alpha\beta}^{\quad AB}\left(  \omega\right)  $ and $e_{A}^{\gamma}.$
Contracting the tangent space index in $R_{\alpha\beta}^{\quad AB}$ with
$e_{A}^{\sigma}$ always removes the $F$ term in $\left(  \ref{40}\right)  $
thanks to $\left(  \ref{16b}\right)  .$There exist only one scalar invariant
in the linear order in curvature%
\begin{equation}
R_{\alpha\beta}^{\quad AB}\left(  \omega\right)  e_{A}^{\alpha}e_{B}^{\beta
}=R\left(  \Gamma\right)  , \label{41}%
\end{equation}
where $R\left(  \Gamma\right)  $ is the usual scalar curvature of 4d manifold
which gives us the Einstein action. Second order invariants in curvature which
are obtained by contracting $R_{\alpha\beta}^{\quad AB}R_{\gamma\delta}^{\quad
CD}$ with four soldering forms $e_{A}e_{B}e_{C}e_{D}$ in all possible
combinations of indices $\alpha\beta\gamma\delta$ give us the space-time
curvature invariants%
\begin{equation}
R^{2}\left(  \Gamma\right)  ,\text{ }R_{\alpha\beta}\left(  \Gamma\right)
R^{\alpha\beta}\left(  \Gamma\right)  ,\text{ }R_{\alpha\beta\gamma\delta
}\left(  \Gamma\right)  R^{\alpha\beta\gamma\delta}\left(  \Gamma\right)  ,
\label{42}%
\end{equation}
and only the contraction of tangent space indices with themselves generate
kinetic terms for $A_{\beta}^{\hat{I}\hat{J}}:$%
\begin{equation}
g^{\alpha\gamma}g^{\beta\delta}R_{\alpha\beta}^{\quad AB}\left(
\omega\right)  R_{\gamma\delta AB}^{\quad}\left(  \omega\right)
=g^{\alpha\gamma}g^{\beta\delta}\left(  F_{\alpha\beta}^{\quad\hat{I}\hat{J}%
}\left(  A\right)  F_{\gamma\delta\hat{I}\hat{J}}\left(  A\right)  \right)
+R_{\alpha\beta\gamma\delta}\left(  \Gamma\right)  R^{\alpha\beta\gamma\delta
}\left(  \Gamma\right)  . \label{43}%
\end{equation}
In this last expression the Yang-Mills kinetic term appears as part of the
gravitational curvature square term.

To summarize, the most general action, up to quadratic order in curvature is
given by%
\begin{align}
I &  =%
{\displaystyle\int}
d^{4}x\sqrt{-g}\left[  \frac{1}{16\pi G}R_{\alpha\beta}^{\quad AB}\left(
\omega\right)  e_{A}^{\alpha}e_{B}^{\beta}-\frac{1}{4}g^{\alpha\gamma}%
g^{\beta\delta}R_{\alpha\beta}^{\quad AB}\left(  \omega\right)  R_{\gamma
\delta AB}^{\quad}\left(  \omega\right)  \right.  \nonumber\\
&  \left.  +R_{\alpha\beta}^{\quad AB}R_{\gamma\delta}^{\quad CD}\left(
ae_{A}^{\alpha}e_{B}^{\beta}e_{C}^{\gamma}e_{D}^{\delta}+be_{A}^{\alpha}%
e_{C}^{\beta}e_{B}^{\gamma}e_{D}^{\delta}+ce_{C}^{\alpha}e_{D}^{\beta}%
e_{A}^{\gamma}e_{B}^{\delta}\right)  \right]  \label{48a}\\
&  =%
{\displaystyle\int}
d^{4}x\sqrt{-g}\left[  \frac{1}{16\pi G}R\left(  \Gamma\right)  +aR^{2}\left(
\Gamma\right)  -bR_{\alpha\beta}\left(  \Gamma\right)  R^{\alpha\beta}\left(
\Gamma\right)  \right.  \nonumber\\
&  \quad\left.  +\left(  c-\frac{1}{4}\right)  R_{\alpha\beta\gamma\delta
}\left(  \Gamma\right)  R^{\alpha\beta\gamma\delta}\left(  \Gamma\right)
-\frac{1}{4}g^{\alpha\gamma}g^{\beta\delta}F_{\alpha\beta}^{\quad\hat{I}%
\hat{J}}\left(  A\right)  F_{\gamma\delta\hat{I}\hat{J}}\left(  A\right)
\right]  \label{48b}%
\end{align}
where $a,b,$ and $c$ are dimensionless constants. We note that it is possible
to avoid the ghost in the graviton propagator by choosing the Gauss-Bonnet
combination of the curvature square terms which corresponds to the choice
$a=\frac{b}{4}=c-\frac{1}{4}.$

The easiest way to understand the above results which showed that the
$SO(1,N-1)$ invariants split into $SO(1,3)$ and $SO(N-4)$ invariants, is to
work in a special gauge. We first split the constraint (\ref{4d}) for
$A=a=1,...,4$ and $A=\hat{I}=5,...N:$%
\begin{align}
0  &  =\partial_{\mu}e_{a}^{\nu}+\omega_{\mu a}^{\quad b}e_{b}^{\nu}%
+\omega_{\mu a}^{\quad\hat{I}}e_{\hat{I}}^{\nu}+\Gamma_{\mu\rho}^{\nu}%
e_{a}^{\rho}\label{part1}\\
0  &  =\partial_{\mu}e_{\hat{I}}^{\nu}+\omega_{\mu\hat{I}}^{\quad a}e_{a}%
^{\nu}+\omega_{\mu\hat{I}}^{\quad\hat{J}}e_{\hat{J}}^{\nu}+\Gamma_{\mu\rho
}^{\nu}e_{\hat{J}}^{\rho} \label{part2}%
\end{align}
The vielbeins $e_{A}^{\mu}$ transform under $SO(1,N-1)$ transformations
according to
\begin{equation}
e_{A}^{\mu}\rightarrow\tilde{e}_{A}^{\mu}=\Lambda_{AB}e^{\mu B}.
\end{equation}
In particular,
\begin{equation}
e_{\hat{I}}^{\mu}\rightarrow\tilde{e}_{\hat{I}}^{\mu}=\Lambda_{\hat{I}a}e^{\mu
a}+\Lambda_{\hat{I}\hat{J}}e^{\mu\hat{J}}.
\end{equation}
The action, by construction, is invariant under $SO(1,N-1)$ rotations. Thus,
it is possible to use the gauge invariance and the freedom in the choice of
gauge parameters $\Lambda_{\hat{I}a}$ to set $e_{\hat{I}}^{\mu}$ to zero%
\begin{equation}
e_{\hat{I}}^{\mu}=0. \label{gauge}%
\end{equation}
This leaves the gauge parameters $\Lambda_{ab}$ and $\Lambda_{\hat{I}\hat{J}}$
arbitrary, corresponding to invariance under the subgroup $SO(1,3)\times
SO(N-4).$ With this gauge choice we see that equation (\ref{part2}) implies%
\begin{equation}
\omega_{\mu\hat{I}}^{\quad a}=0,
\end{equation}
assuming that $e_{a}^{\mu}$ is invertible. The remaining equation
(\ref{part1}) can now be solved to give the usual expression for $\omega_{\mu
a}^{\quad b}$ in terms of $e_{a}^{\mu}$ and its derivative. In this special
gauge $\omega_{\mu\hat{I}}^{\quad\hat{J}}=A_{\mu\hat{I}}^{\quad\hat{J}}$ and
\begin{equation}
R_{\mu\nu}^{\quad a\hat{I}}=0,
\end{equation}
while nonvanishing components of the curvature $R_{\mu\nu}^{\quad ab}$ and
$R_{\mu\nu}^{\quad\hat{I}\hat{J}}$ are responsible for the gravity and gauge
fields respectively.

Thus, the gauge groups can be considered as subgroup of the Lorentz group of a
higher dimensional tangent space. The connections $A_{\alpha}^{\hat{I}\hat{J}%
}$ transform under $SO\left(  N-4\right)  $ rotations in a subspace orthogonal
to the space spanned by coordinate tangent vectors. The gauge fields come
unified with gravity within $SO\left(  1,N-1\right)  $ Lorentz group. In case
$N=5,$ the connection $A_{\alpha}^{55}$ vanishes and there are no extra gauge
fields in addition to gravity in agreement with \cite{CMtangent}. For $N=6$
the connection $A_{\alpha}^{56}$ is a Maxwell field and the local gauge group
$SO\left(  2\right)  $ is obviously isomorphic to the $U\left(  1\right)  $
group of electrodynamics. Thus, electromagnetism is unified with gravity in
$SO\left(  1,5\right)  $ tangent space group. The realistic group which can
allow us to unify all known interactions is $SO\left(  1,13\right)  .$ In this
case in addition to gravity, the theory describes $45$ dynamical gauge fields
$A_{\alpha}^{\hat{I}\hat{J}}$ which transform under $SO\left(  10\right)  $
group. However, if one wishes to also include family unification, then a
larger group such as $SO\left(  1,21\right)  $ would be needed.

\section{Fermions}

The matter content of the theory, described by fermions, must be in the
fundamental spinor representation of the corresponding Lorentz group
$SO\left(  1,N-1\right)  .$ At this point, it is useful to make a scan of
possible unification groups by considering various dimensions of the tangent
space in four dimensional manifold.

When the tangent space have only one extra dimension compared to the dimension
of space-time the tangent group is the de Sitter group $SO\left(  1,4\right)
$. In this case $\omega_{\mu}^{\hat{5}\hat{5}}=0$ because $\omega_{\mu}^{AB}$
is skew-symmetric in tangent indices and there is no gauge group in addition
to the gravity. The spinors are defined in the $SO(1,4)$ tangent space, where
neither the Weyl nor the Majorana condition could be imposed. By changing the
signature of $SO\left(  1,4\right)  $ to $SO\left(  2,3\right)  $ the Majorana
condition could be imposed. The $SO\left(  2,3\right)  $ case is completely
identical to General Relativity with $SO(1,3)$ tangent group \cite{Who} .

For $N=6$ the gauge group is $SO(2)$ and it describes the Maxwell field. The
spinors are in the $SO\left(  1,5\right)  $ tangent space, where a
symplectic-Majorana or Weyl condition can be imposed. The Clifford algebra is
then $Cl\left(  1,5\right)  =\mathbb{H}\left(  4\right)  $ and the spinor is
of dimension $8.$ It reduces to two independent spinors when the
symplectic-Majorana or Weyl condition is imposed, which are equivalent to a
Dirac spinor, or a pair of Majorana spinors with respect to $SO(1,3).$

In a seven dimensional tangent space $\left(  N=7\right)  $ the gauge group is
$SO\left(  3\right)  $, which is locally isomorphic to $SU\left(  2\right)  .$
The Clifford algebra of the $SO\left(  1,6\right)  $ tangent group is
$Cl\left(  1,6\right)  =\mathbb{C}\left(  8\right)  $ and the spinor is of
dimension $8$. No further conditions can be imposed in this case to reduce the
number of independent components. The spinor is of the form $\psi_{\alpha i}$
with $i=1,2$ in the spinor representation of $SO\left(  3\right)  $ and it is
a Dirac spinor with respect to the index $\alpha.$

When $N=8,$ the gauge group is $SO\left(  4\right)  $ and the tangent group is
$SO\left(  1,7\right)  .$ The Clifford algebra for this tangent group is
$Cl\left(  1,7\right)  =%
\mathbb{R}
\left(  16\right)  $ and the spinor is of dimension $16.$ It can be subject to
the Weyl condition, thus, reducing the number of independent components to
$8.$ Since $SO\left(  4\right)  $ is locally isomorphic to $SU\left(
2\right)  \times SU\left(  2\right)  $ the spinor is of the form $\psi_{\alpha
i}$ and $\psi_{\alpha i^{\prime}}$ where $i=1,2$ and $i^{\prime}=1,2$ are in
the spinor representations of the two $SU\left(  2\right)  .$

Continuing this consideration to higher $N$ we find that the smallest rotation
group that has $SU(3)\times SU(2)\times U(1)$ gauge group of the Standard
Model as a subgroup is $SO(10)$ and a good candidate for the realistic model
which unifies gravity with gauge interactions is
\begin{equation}
\mathcal{G}=SO(1,13)\label{46}%
\end{equation}
local symmetry group of the tangent space in the four dimensional manifold$.$
A spinor $\psi_{\widehat{\alpha}}$ in the fundamental representation of
$SO(1,13)$ has $2^{7}=128$ components on which one can impose a Weyl condition%
\begin{equation}
\left(  \Gamma_{15}\right)  _{_{\widehat{\alpha}}}^{_{\widehat{\beta}}}%
\psi_{\widehat{\beta}}=\psi_{\widehat{\alpha}}\label{47}%
\end{equation}
where $\Gamma_{15}=\Gamma_{0}\Gamma_{1}\cdots\Gamma_{13}$ satisfies $\left(
\Gamma_{15}\right)  ^{2}=1$ and $\Gamma_{0},\Gamma_{1},...,\Gamma_{13}$ are
fourteen $2^{7}\times2^{7}$ gamma matrices that satisfy the Clifford algebra
$Cl\left(  1,13\right)  $. The Weyl condition reduces the number of
independent components of the spinor to $\frac{1}{2}\left(  128\right)  =64.$
This corresponds to a Dirac $SO\left(  1,3\right)  $ spinor in the
$16_{s}+\overline{16}_{s}$ representation of $SO(10).$ The $64$ independent
component spinor describes $32$ two components Weyl fermions. Thus, the number
of fermions in the fundamental spinor representation of $SO\left(
1,13\right)  $ is twice more than in the Standard Model, where one family
contains only $16$ Weyl fermions. A Majorana mass term that couples $16_{s}$
to $\overline{16}_{s}$ vanishes identically in this case and one has to appeal
to some mechanism to make the mirror fermions very heavy to avoid conflict
with experiments. Such mechanism could be borrowed from the study of
topological superconductors, where only the mirror fermions $\overline{16}%
_{s}$ acquire mass without breaking the $SO\left(  10\right)  $ symmetry
\cite{Kitaev}, \cite{Wen}, \cite{Zee}. Consider the coupling,
\begin{equation}
\overline{\psi}_{\widehat{\alpha}}\left(  \lambda\left(  \Gamma^{A}\right)
_{\widehat{\alpha}}^{\widehat{\beta}}H_{A}+\lambda^{\prime}\left(
\Gamma^{ABCDE}\right)  _{\widehat{\alpha}}^{\widehat{\beta}}H_{ABCDE}\right)
\psi_{\widehat{\beta}}%
\end{equation}
where $H_{A}$ and $H_{ABCDE}$ are Higgs fields whose kinetic terms are
completely antisymmetrized%
\begin{align}
F_{AB} &  =D_{A}H_{B}-D_{B}H_{A}\\
F_{ABCDEF} &  =5\left[  D_{A}\right.  H_{BCD\left.  E\right]  }%
\end{align}
where $D_{A}=e_{A}^{\mu}D_{\mu}.$ If $H_{A}$ and $H_{ABCDE}$ develop vev
\begin{align}
\left\langle H_{A}\right\rangle  &  =\delta_{A}^{I}\phi_{I}\\
\left\langle H_{ABCDE}\right\rangle  &  =\frac{1}{4!}\delta_{\left[
ABCD\right.  }^{abcd}\delta_{\left.  E\right]  }^{I}\epsilon_{abcd}\phi_{I}%
\end{align}
The field $\phi_{I}$ will then couple to the $\chi=16_{s}$ and $\chi^{\prime
}=\overline{16}_{s}$ with couplings $\lambda_{10}=\lambda-\lambda^{\prime}$
and and $\lambda_{10}^{\prime}=\lambda+\lambda^{\prime}$%
\begin{equation}
\left(  \lambda_{10}\overline{\chi}\Gamma^{I}\chi+\lambda_{10}^{\prime
}\overline{\chi}^{\prime}\Gamma^{I}\chi^{\prime}\right)  \phi_{I}%
\end{equation}
It is then assumed that the couplings $\lambda_{10}\ll1$ and  $\lambda
_{10}^{\prime}\sim1$ so that the fermions $\chi$ would see $\phi_{I}$ as a
Higgs field and the mirror fermions $\chi^{\prime}$ would see $\phi_{I}$ as
collection of wildly fluctuating scalar fields which could be replaced with
constants and thus making the $\overline{16}_{s}$ super-heavy \cite{Zee}.
Further breaking of $SO\left(  10\right)  $ is then achieved by the usual
Higgs mechanism using appropriate scalar fields \cite{fakir}. One can go
further and allow for family unification \ by considering the group $SO\left(
1,21\right)  $ which has the same spinorial properties as $SO\left(
1,13\right)  $ as the dimensions differ by $8.$ Majorana mass terms that
couples the $16_{s}$ and $\overline{16}_{s}$ vanish. The grand unified group
in this case is $SO\left(  18\right)  $ and a full analysis of such model is
given in \cite{Zee} where it is argued that it is possible to obtain three
families of $\overline{5}+10$ of $SU\left(  5\right)  $ and where all the
extra matter acquires mass. The symmetry breaking takes the route
\begin{equation}
SO\left(  18\right)  \rightarrow U\left(  9\right)  \rightarrow SU\left(
5\right)  \times SU\left(  4\right)  \rightarrow SU\left(  5\right)  \times
SO\left(  5\right)
\end{equation}
The last stages of breaking are done through the usual Higgs mechanism.

\section{Conclusions}

We have shown that one can unify gauge interactions with gravity by
considering higher dimensional tangent groups in a four dimensional
space-time. The gauged Lorentz group of the tangent space describes
simultaneously the symmetry groups of gravity and gauge interactions, provided
a metricity condition is satisfied. The spin-connections of the higher
dimensional tangent space fully incorporate information on the affine
connection of space-time as well as the gauge fields. Those connections which
are responsible for gravity are \textquotedblleft composite\textquotedblright%
\ because they satisfy extra constraints which allow to express them in terms
of the derivatives of the vielbeins. On the other hand the spin-connections
responsible for gauge interactions do not obey any constraints and hence are
independent. The complete geometric unification of gravity and gauge
interactions is realized by writing the action of the theory just in terms of
curvature invariants of the tangent group which contains the Yang-Mills action
for gauge fields.

A realistic group which unifies gravity with gauge interactions and contains
the Standard Model is $SO\left(  1,13\right)  \,\ $in a fourteen dimensional
tangent space. It corresponds to $SO\left(  10\right)  $ grand unified theory
concerning the gauge fields content, however, it has double the number of
fermions in the form of $16_{s}+\overline{16}_{s}$. It is not easy to decouple
the mirror fermions by giving them very heavy masses via Brout-Englert-Higgs
mechanism. Instead, this could be done by appealing to a mechanism used for
topological superconductors, where the $\overline{16}_{s}$ could be made very
heavy. One can go further and unify the three families by considering
$SO\left(  1,21\right)  $ instead of $SO\left(  1,13\right)  $ with $SO\left(
18\right)  $ grand unification group as argued in \cite{Zee}. Since the Dirac
operator plays a fundamental role in this setting, it is natural to look for
connections between this construction and that of noncommutative geometry. In
addition, the need to add Higgs scalar fields suggests that a total
unification of gravity, gauge and Higgs fields within one geometrical setting,
should be possible by replacing the continuous four-dimensional manifold by a
noncommutative space which has both discrete and continuous structures
\cite{NCG}. This possibility and others will be the subject of future investigations.

\textit{Notes added}

\begin{itemize}
\item After this paper was submitted we were informed by R. Percacci of his
work in references \cite{Precacci1},\cite{Percacci3} \cite{Percacci2}. In
reference \cite{Precacci1} a $GL(4,R)$ model is considered with torsion and a
connection with non-metricity. In reference \cite{Percacci3} this is
generalized to $GL\left(  N,R\right)  $ broken spontaneously to $O\left(
1,N-1\right)  .$ In reference \cite{Percacci2} the issue of chiral fermions in
a gauged $SO\left(  3,11\right)  $ model broken to $SO\left(  3,1\right)
\times SO\left(  10\right)  $ where the Majorana-Weyl condition is imposed to
avoid mirror fermions. This model does suffer from the presence of ghosts for
scalar Higgs fields and whenever the Minkowski metric is used an odd number of
times. Although the methods in these works are similar to the ones presented
here, there is little overlap.

\item Michel Dubois-Violette, communicated to us the following. In 1970, R.
Greene has proved that a 4-dimensional Lorentzian manifold admits locally an
isometric smooth free embedding in Minkowski space $M(1,13)$ \cite{Greene}.
There is a similar result proved the same year for the Euclidean signature in
M.L. Gromov and V.A. Rokhlin \cite{Gromow}. This means that one can include an
arbitrary deformation of the four-manifold in the same flat space and
eventually expect to \textit{quantize} space-time in the fixed Minkowski space
$M(1,13).$

\item Latham Boyle pointed to us the relevance of the work on topological
insulators and superconductors to give mass to mirror fermions as in
references \cite{Wen} and \cite{Kitaev}.
\end{itemize}

\begin{acknowledgement}
A.H.C would like to thank the Pauli Center at the Institute for Theoretical
Physics, ETH, Zurich, Switzerland and the Arnold Sommerfeld Institute for
Theoretical Physics at the Ludwig Maximilians University, Munich, Germany for
their hospitality where this research was done. His work is also supported in
part by the National Science Foundation Grant No. Phys-1202671 and
Phys-1518371. The work of V.M is supported by TRR 33 \textquotedblleft The
Dark Universe\textquotedblright\ and the Cluster of Excellence EXC 153
\textquotedblleft Origin and Structure of the Universe\textquotedblright.
\end{acknowledgement}

\end{document}